\documentclass[runningheads,fleqn]{llncs}

\usepackage[utf8]{inputenc}
\usepackage[T1]{fontenc}
\usepackage{amsmath}
\usepackage{latexsym}
\usepackage{subcaption}
\usepackage{graphicx}
\usepackage{xspace}
\usepackage{tikz}

\newcommand{\tlaplus}{TLA\textsuperscript{+}\xspace}
\newcommand{\tlaps}{\textsc{tlaps}\xspace}
\newcommand{\tlc}{\textsc{tlc}\xspace}
\newcommand{\apalache}{\textsc{apalache}\xspace}

\newcommand{\kw}[1]{\textsc{#1}}
\newcommand{\deq}{\mathrel{\smash{\stackrel{\scriptscriptstyle\Delta}{=}}}}
\newcommand{\seq}[1]{\langle #1 \rangle}
\newcommand{\str}[1]{\text{\textsf{\upshape ``#1''}}}
\newcommand{\ps}[2]{\langle #1 \rangle #2}
\let\implies\undefined{}    
\newcommand{\implies}{\Rightarrow}
\newcommand{\biimplies}{\ensuremath{\Leftrightarrow}}
\newcommand{\A}{\forall}
\newcommand{\E}{\exists}
\newcommand{\alw}{\Box}

\newcommand{\an}[2]{\langle #1 \rangle_{#2}}
\newcommand{\sq}[2]{[#1]_{#2}}
\newcommand{\WF}{\textrm{WF}}

\newcommand{\fun}{\rightarrow}

\newenvironment{conj}{\begin{array}[t]{@{\mbox{$\land$\ }}l@{}}}{\end{array}}
\newenvironment{disj}{\begin{array}[t]{@{\mbox{$\lor$\ }}l@{}}}{\end{array}}
\newenvironment{noj}{\begin{array}[t]{@{}l@{}}}{\end{array}}
\newenvironment{noj2}{\begin{array}[t]{@{}l@{\ }l@{}}}{\end{array}}

\makeatletter

\newlength{\@charwidth}
\newlength{\boxrulewd}
\newlength{\boxindent}
\newlength{\boxlineht}
\newcommand{\boxsep}{\@charwidth}
\newlength{\boxruleht}
\newlength{\boxruledp}
\newcommand{\@computerule}{%
  \setlength{\boxruleht}{.5ex}%
  \setlength{\boxruledp}{-\boxruleht}%
  \addtolength{\boxruledp}{\boxrulewd}}
\newcommand{\boxrule}{\leaders\hrule height \boxruleht depth \boxruledp
                      \hfill\mbox{}}

\newcommand{\@nmlineraise}{\if@botline -\boxrulewd
    \else-\boxlineht\fi}
\newcommand{\nameline}[1]{\hspace*{-\boxsep}%
    \raisebox{\@nmlineraise}[0pt][0pt]{\rule[.5ex]{\boxrulewd
               }{\boxlineht}}%
    \boxrule
    #1\boxrule
    \raisebox{\@nmlineraise}[0pt][0pt]{\rule[.5ex]{\boxrulewd
               }{\boxlineht}}\hspace{-\boxsep}\vspace{.5\baselineskip}}
\newif\if@botline

\newcommand{\topbar}[1]{\tlanameline{\ 
     {\sc module} {\it #1\/} }\par}

\newenvironment{nomodule}[1][.92]{\begin{minipage}{#1\linewidth}\begin{trivlist}\raggedright\@computerule
            \item[]\it
        }{\par\end{trivlist}\end{minipage}}

\newcommand{\tlanameline}[1]{\hspace*{-\boxsep}%
    \raisebox{\@nmlineraise}[0pt][0pt]{\rule[.5ex]{\boxrulewd
               }{\boxlineht}}%
    \boxrule
    #1\boxrule
    \raisebox{\@nmlineraise}[0pt][0pt]{\rule[.5ex]{\boxrulewd
               }{\boxlineht}}\hspace{-\boxsep}%
          \vspace{.2em}
               }

\newcommand{\midbar}{\par\hspace{-\boxsep}%
    \raisebox{-.5\boxlineht}[0pt][0pt]{\rule[.5ex]{\boxrulewd
               }{\boxlineht}}%
     \boxrule
    \raisebox{-.5\boxlineht}[0pt][0pt]{\rule[.5ex]{\boxrulewd
               }{\boxlineht}}\hspace{-\boxsep}%
         \vspace{-.2em}
                \par}

\newcommand{\bottombar}{\hspace{-\boxsep}%
    \raisebox{-\boxrulewd}[0pt][0pt]{\rule[.5ex]{\boxrulewd
               }{\boxlineht}}%
     \boxrule
    \raisebox{-\boxrulewd}[0pt][0pt]{\rule[.5ex]{\boxrulewd
               }{\boxlineht}}\hspace{-\boxsep}\vspace{0pt}}

\newcommand{\xtrivlist}{\begin{list}{}{%
       \leftmargin=1em%
       \labelwidth=0pt%
       \labelsep=0pt%
       \parsep=0pt%
       \topsep=0pt%
       \partopsep=0pt%
       \itemsep=0pt}}
\def\endxtrivlist{\end{list}}

\let\@@startfield\@startfield
\let\@@stopfield\@stopfield

\def\tla@tabcr{\@stopline\@startline\ignorespaces}

\def\xtabbing{\lineskip \z@\let\>\@rtab\let\<\@ltab\let\=\@settab
     \def\@startfield{\@@startfield\(}
     \def\@stopfield{\)\@@stopfield}
     \let\+\@tabplus
     \let\-\@tabminus
     \let\`\@tabrj\let\'\@tablab
     \let\\=\tla@tabcr
     \global\@hightab\@firsttab
     \global\@nxttabmar\@firsttab
     \dimen\@firsttab\@totalleftmargin
     \global\@tabpush0 \global\@rjfieldfalse
     \xtrivlist \item[]\if@minipage\else\vskip\parskip\fi
     \setbox\@tabfbox\hbox{\rlap{\indent\hskip\@totalleftmargin
       \the\everypar}}\def\@itemfudge{\box\@tabfbox}\@startline\ignorespaces}

\def\endxtabbing{\@stopline\ifnum\@tabpush > 0 \@badpoptabs \fi\endxtrivlist}

\newcommand{\EXTENDS}{\mbox{\sc extends}}
\newcommand{\INSTANCE}{\mbox{\sc instance}}
\newcommand{\THEOREM}{\mbox{\sc theorem}}
\newcommand{\LEMMA}{\mbox{\sc lemma}}

\newcommand{\ASSUME}{\mbox{\sc assume}}
\newcommand{\PROVE}{\mbox{\sc prove}}

\newcommand{\VARIABLES}{\mbox{\sc variables}}
\newcommand{\CONSTANT}{\mbox{\sc constant}}

\newcommand{\BOOLEAN}{\mbox{\sc boolean}}

\newcommand{\UNCHANGED}{\mbox{\sc unchanged}}
\newcommand{\EXCEPT}{\mbox{\sc except}}

\newcommand{\ENABLED}{\mbox{\sc enabled}}

\newcommand{\SUBSET}{\mbox{\sc subset}}
\newcommand{\BY}{\mbox{\sc by}}

\newcommand{\DEF}{\mbox{\sc def}}
\newcommand{\QED}{\mbox{\sc qed}}
\newcommand{\TRUE}{\kw{true}}
\newcommand{\FALSE}{\kw{false}}
\newcommand{\DEFINE}{\kw{define}}

\newcommand{\NEW}{\kw{new}}

\newcommand{\PTL}{\kw{ptl}}

\newcounter{abr@ctr}
\newcommand{\abr@c}{\c@abr@ctr\advance\c@abr@ctr\@ne}

\ifx\documentclass\undefined
\else
  \DeclareSymbolFont{tlaitalics}{\encodingdefault}{cmr}{m}{it}
  \let\itfam\symtlaitalics
\fi

\newcommand{\noTeXmath}{%
\c@abr@ctr=\itfam
\multiply\c@abr@ctr"100\relax
\advance\c@abr@ctr "7061\relax
\mathcode`a=\abr@c\mathcode`b=\abr@c\mathcode`c=\abr@c\mathcode`d=\abr@c
\mathcode`e=\abr@c\mathcode`f=\abr@c\mathcode`g=\abr@c\mathcode`h=\abr@c
\mathcode`i=\abr@c\mathcode`j=\abr@c\mathcode`k=\abr@c\mathcode`l=\abr@c
\mathcode`m=\abr@c\mathcode`n=\abr@c\mathcode`o=\abr@c\mathcode`p=\abr@c
\mathcode`q=\abr@c\mathcode`r=\abr@c\mathcode`s=\abr@c\mathcode`t=\abr@c
\mathcode`u=\abr@c\mathcode`v=\abr@c\mathcode`w=\abr@c\mathcode`x=\abr@c
\mathcode`y=\abr@c\mathcode`z=\abr@c
\c@abr@ctr=\itfam
\multiply\c@abr@ctr"100\relax
\advance\c@abr@ctr "7041\relax
\mathcode`A=\abr@c\mathcode`B=\abr@c\mathcode`C=\abr@c\mathcode`D=\abr@c
\mathcode`E=\abr@c\mathcode`F=\abr@c\mathcode`G=\abr@c\mathcode`H=\abr@c
\mathcode`I=\abr@c\mathcode`J=\abr@c\mathcode`K=\abr@c\mathcode`L=\abr@c
\mathcode`M=\abr@c\mathcode`N=\abr@c\mathcode`O=\abr@c\mathcode`P=\abr@c
\mathcode`Q=\abr@c\mathcode`R=\abr@c\mathcode`S=\abr@c\mathcode`T=\abr@c
\mathcode`U=\abr@c\mathcode`V=\abr@c\mathcode`W=\abr@c\mathcode`X=\abr@c
\mathcode`Y=\abr@c\mathcode`Z=\abr@c}

\makeatother
\noTeXmath

\begin{document}

\title{Specification and Verification With the \tlaplus Trifecta: TLC, Apalache, and TLAPS%
  \thanks{Support by the Inria-Microsoft Research Joint Centre and 
    Interchain Foundation, Switzerland is gratefully acknowledged.}}
\titlerunning{Specification and Verification With the \tlaplus Trifecta}

\author{%
  Igor Konnov\inst{1} 
  \and
  Markus Kuppe\inst{2} 
  \and
  Stephan Merz\inst{3} 
}

\authorrunning{I.\ Konnov, M.\ Kuppe, and S.\ Merz}

\institute{%
  Informal Systems, Vienna, Austria \and
  Microsoft Research \and
  University of Lorraine, CNRS, Inria, LORIA, Nancy, France
}

\maketitle              
\begin{abstract}
  Using an algorithm due to Safra for distributed termination detection as a
  running example, we present the main tools for verifying specifications
  written in \tlaplus. Examining their complementary strengths and weaknesses,
  we suggest a workflow that supports different types of analysis and that can
  be adapted to the desired degree of confidence.
  
\end{abstract}

\section{Introduction}
\label{sec:intro}

\tlaplus~\cite{lamport:specifying} is a formal language for specifying systems,
in particular concurrent and distributed algorithms, at a high level of
abstraction. The foundations of \tlaplus are classical Zermelo-Fraenkel set
theory with choice for representing the data structures on which the algorithm
operates, and the Temporal Logic of Actions (TLA), a variant of linear-time
temporal logic, for describing executions of the algorithm.

Most users write \tlaplus specifications using one of the two existing IDEs
(integrated development environments): the \tlaplus
Toolbox~\cite{DBLP:journals/corr/abs-1912-10633}, a standalone Eclipse
application, and a Visual Studio Code Extension, which can be run in a standard
Web browser. To different degrees, both IDEs also integrate the three main tools
for verifying \tlaplus specifications: the explicit-state model checker
\tlc~\cite{yu:model-checking}, the symbolic SMT-based model checker
\apalache~\cite{konnov:apalache}, and the \tlaplus Proof System
\tlaps~\cite{cousineau:tla-proofs}, an interactive proof assistant.

In this paper, we use a non-trivial algorithm for detecting termination of an
asychronous distributed system~\cite{dijkstra:ewd998} as a running example for
presenting the three tools. Based on their complementary strengths and
weaknesses, we suggest a workflow that can serve as a guideline for analyzing
different kinds of properties of a \tlaplus specification, to different degrees
of confidence. This is the first paper that applies all three tools to a
common specification and identifies their complementary qualities. We hope that
future \tlaplus users will find our presentation useful for applying these
tools to their own specifications. All of our \tlaplus modules and ancillary
files for running the tools are available online~\cite{MerzKuppeKonnov2021}.

\paragraph{Outline of the paper.}

Section~\ref{sec:td} presents a high-level specification of the problem of
termination detection written as a \tlaplus state machine, whose properties are
verified in Section~\ref{sec:atd-verification}. Safra's algorithm is informally
presented in Section~\ref{sec:ewd998}. Section~\ref{sec:ewd998-analysis}
describes different approaches for verifying the properties of the algorithm
using model checking and theorem proving, including checking that the algorithm
refines the high-level specification introduced earlier. Finally,
Section~\ref{sec:conclusion} concludes the paper and outlines ideas for future
work.

\section{Specifying Termination Detection}
\label{sec:td}

Before presenting Safra's algorithm for detecting termination of processes on a
ring, we formally state the problem to be solved. Although this is not strictly
required for using \tlaplus, it will allow us to succinctly state correctness of
the algorithm in terms of refinement, and to introduce the \tlaplus tools on a
small specification.

\begin{figure}[tb]
  \begin{subfigure}[b]{0.33\linewidth}
    \centering
    \begin{tikzpicture}[scale=.7]
      \draw (0,0) circle (2cm) ;
      \draw[fill=white] (0,2) circle (2.5mm) ;
      \draw[fill=white] (0,2) circle (1.7mm) ;
      \draw (0,1.4) node (0) {0} ;
      \draw[fill=white] (2,0) circle (2.5mm) ;
      \draw[fill=white] (2,0) circle (1.7mm) ;
      \draw (1.4,0) node (3) {3} ;
      \draw[fill=white] (0,-2) circle (2.5mm) ;
      \draw[fill=white] (0,-2) circle (1.7mm) ;
      \draw (0,-1.4) node (2) {2} ;
      \draw[fill=white] (-2,0) circle (2.5mm) ;
      \draw (-1.4,0) node (1) {1} ;
    \end{tikzpicture}
    \caption{Visual representation.}
    \label{fig:ring-snapshot}
  \end{subfigure}
  \hfill
  \begin{subfigure}[b]{0.65\linewidth}
    \centering
    \begin{nomodule}
      \topbar{AsynchronousTerminationDetection}
      $\EXTENDS\ \ Naturals$\\
      $\CONSTANT\ \ N$\\
      $\ASSUME\ \ NAssumption\ \deq\ N \in Nat \setminus \{0\}$\\[.5ex]
      $Node\ \deq\ 0 \,..\, N\!-\!1$\\[.5ex]
      $\VARIABLES\ \ active, pending, termDetect$\\[.5ex]
      $TypeOK\ \deq\ 
      \begin{conj}
        active \in [Node \fun \BOOLEAN]\\
        pending \in [Node \fun Nat]\\
        termDetect \in \BOOLEAN
      \end{conj}$\\
      $vars\ \deq\ \seq{active, pending, termDetect}$\\
      $terminated\ \deq$\\
      \hspace*{1em}$\A n \in Node : \lnot active[n] \land pending[n] = 0$
      \midbar
    \end{nomodule}

    \caption{\tlaplus representation of the state space.}
    \label{fig:ring-tla}
  \end{subfigure}
  \caption{The ring of nodes and its \tlaplus representation.}
  \label{fig:ring}
\end{figure}
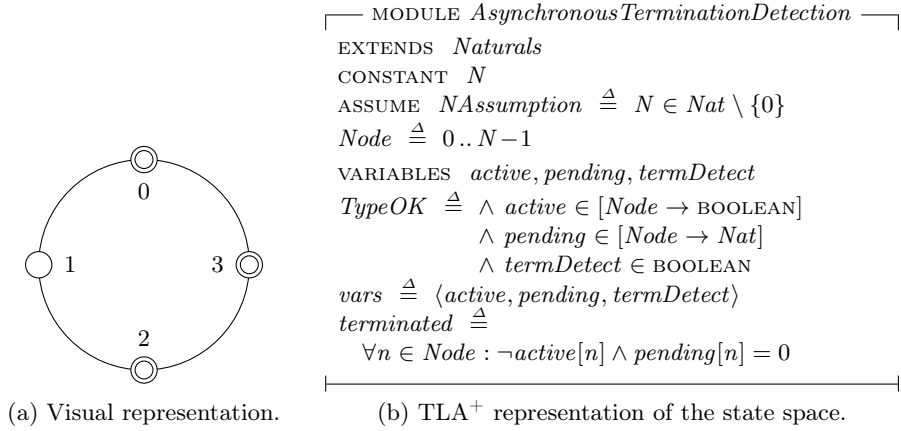

As illustrated in Fig.~\ref{fig:ring-snapshot}, we assume $N$ nodes that perform
some distributed computation. Each node can be active (indicated by a double
circle) or inactive (single circle). When a node is active, it may perform some
local computation, and it can send messages to other nodes. Messages can be
understood as carrying tasks to be performed by the receiver: an inactive node
can still receive messages and will then become active. The purpose of the
algorithm is to detect whether all nodes are inactive. Note that the ring
structure indicated in Fig.~\ref{fig:ring-snapshot} is unimportant for the
statement of the problem: it will become relevant for the termination detection
algorithm to be introduced later.

Figure~\ref{fig:ring-tla} contains the initial part of a \tlaplus specification
that formally states the problem of termination detection. Specifications appear
in \tlaplus modules that contain declarations of constant and variable
parameters, statements of assumptions and theorems, but mainly contain operator
definitions of the form $Op(args) \deq expr$; if the operator does not take
arguments, the empty pair of brackets is omitted.

Whereas constant parameters are interpreted by fixed (state-independent) values,
variable parameters correspond to state variables that represent the evolution
of a system: a \emph{state} assigns values to variables. \tlaplus expressions
are classified in four levels: \emph{constant formulas}\footnote{In \tlaplus
  jargon, the term ``formula'' denotes any expression, not necessarily Boolean.}
do not contain any variables, thus their value is the same at all states during
the execution of a system. \emph{State formulas} may contain constants and
unprimed variables, and such formulas are evaluated at individual states. \emph{Action
  formulas} may additionally contain primed variables; they are evaluated over pairs
$(s,t)$ of states: an unprimed variable~$v$ denotes the value of~$v$ in~$s$
whereas the primed variable~$v'$ denotes the value of~$v$ in~$t$. Finally,
\emph{temporal formulas} involve operators of temporal logic such as~$\Box$ (\emph{always}) 
and~$\Diamond$ (\emph{eventually}), and they are evaluated over (infinite) 
sequences of states. By abuse of language, we will sometimes say that a module
defines a state predicate rather than that it defines an operator representing
a Boolean state formula, and similarly for the other levels.

\tlaplus modules form a hierarchy by extending and instantiating other
modules. Module $AsynchronousTerminationDetection$ extends module $Naturals$
from the standard library in order to import the definition of the set $Nat$ of
natural numbers and standard arithmetic operations. It declares a constant
parameter~$N$ and defines the constant $Node$ as the interval of integers
between $0$ and $N-1$. It also declares variable parameters $active$, $pending$,
and $termDetect$, intended to represent the status of each node (active or
inactive), the number of pending messages at each node, and whether termination
has been detected or not. \tlaplus is untyped, but it is good engineering
practice to document the expected types of constants and variables using a formula
of the shape $v \in S$ where $S$ is a set that denotes the type of~$v$. Thus,
assumption $NAssumption$ states that~$N$ must be a non-zero natural number,
and the state predicate $TypeOK$ indicates the expected types of the
variables:\footnote{In \tlaplus, long conjunctions and disjunctions are
  conventionally written as lists whose ``bullets'' are the logical operator,
  with nesting indicated by indentation.} $active$ and $pending$ are Boolean and
natural-number valued functions over $Node$, whereas $termDetect$ is a Boolean
value. The formal status of the two typing predicates is quite different: an
instance of the module that does not satisfy $NAssumption$ is considered
illegal, whereas $TypeOK$ just defines a predicate. (We will
later check that the predicate indeed holds at every state during any execution.)
Finally, the module declares a predicate $terminated$ that is true at states in
which the system has globally terminated: all nodes are inactive, and no message
is pending at any node.

\begin{figure}[tb]
  \begin{nomodule}[1]
  \begin{minipage}[t]{0.55\linewidth}
    $Init\ \deq$\\
    \hspace*{1em}$\begin{conj}
      active \in [Node \fun \BOOLEAN]\\
      pending = [n \in Node \mapsto 0]\\
      termDetect \in \{\FALSE, terminated\}
    \end{conj}$\\
    $Terminate(i)\ \deq$\\
    \hspace*{1em}$\begin{conj}
      active[i]\\
      active' = [active\ \EXCEPT\ ![i] = \FALSE]\\
      pending' = pending\\
      termDetect' \in \{termDetect, terminated'\}
    \end{conj}$\\
    $SendMsg(i,j)\ \deq$\\
    \hspace*{1em}$\begin{conj}
      active[i]\\
      pending' = [pending\ \EXCEPT\ ![j] = @ + 1]\\
      \UNCHANGED\ \seq{active, termDetect}
    \end{conj}$\\
    $RcvMsg(i)\ \deq$\\
    \hspace*{1em}$\begin{conj}
      pending[i] > 0\\
      active' = [active\ \EXCEPT\ ![i] = \TRUE]\\
      pending' = [pending\ \EXCEPT\ ![i] = @ - 1]\\
      \UNCHANGED\ termDetect
    \end{conj}$
  \end{minipage}
  \hfill
  \begin{minipage}[t]{0.44\linewidth}
    $DetectTermination\ \deq$\\
    \hspace*{1em}$\begin{conj}
      terminated\\
      termDetect' = \TRUE\\
      \UNCHANGED\ \seq{active, pending}
    \end{conj}$\\
    $Next\ \deq$\\
    \hspace*{1em}$\begin{disj}
      \E i \in Node : RcvMsg(i)\\
      \E i \in Node : Terminate(i)\\
      \E i,j \in Node : SendMsg(i,j)\\
      DetectTermination
    \end{disj}$\\
    $Spec\ \deq$\\
    \hspace*{1em}$\begin{conj}
      Init \land \alw\sq{Next}{vars}\\
      \WF_{vars}(DetectTermination)
    \end{conj}$
  \end{minipage}\\
  \bottombar
  \end{nomodule}
  \caption{End of module \textit{AsynchronousTerminationDetection}.}
  \label{fig:atd-tla}
\end{figure}

The part of the specification shown in Fig.~\ref{fig:ring-tla} corresponds to
the static model of the system. Module $AsynchronousTerminationDetection$
continues as shown in Fig.~\ref{fig:atd-tla} with the specification of a
transition system that abstractly represents the problem of distributed
termination detection. The initial condition $Init$ expresses that the variable
$active$ can take any type-correct value and that no messages are pending. As
for $termDetect$, it will usually be $\FALSE$ initially, but it could be $\TRUE$
in case the predicate $terminated$ holds.\footnote{Remember that the purpose of
  this specification is not to describe a specific termination detection
  algorithm, but rather an abstract transition system meant to represent any
  such algorithm.} The transition formulas $Terminate(i)$, $SendMsg(i,j)$,
$RcvMsg(i)$, and $DetectTermination$ describe the allowed transitions. For
example, $Terminate(i)$ describes local termination of node~$i$. The action is
possible if node~$i$ is active, the value of variable $active$ after the 
transition is similar to its previous value, except that $active[i]$ is $\FALSE$,
and variable $pending$ is left unchanged. The variable $termDetect$ may also 
remain unchanged, but a
clever termination algorithm may set it to $\TRUE$ provided the predicate
$terminated$ has become true with the termination of node~$i$. Note that updates
of function-valued variables are expressed using $\EXCEPT$ clauses, which
describe for which arguments the function is set to a new value. On the
right-hand side of such a clause, the symbol~$@$ refers to the old value of the
function at the argument that is updated. The definitions of the remaining actions
are similar.

The action $Next$ is defined as the disjunction of the actions introduced
before, and the temporal formula $Spec$ corresponds to the overall specification
of the system behavior. It has the standard form
\[
  Init \land \alw\sq{Next}{vars} \land L
\]
of a \tlaplus specification, asserting that executions must start in a state
satisfying the initial condition and that all transitions must either correspond
to a system transition (described by formula $Next$) or leave the variables
$vars$ unchanged. The supplementary condition $L$ is usually a conjunction of
fairness conditions; in our example, we require that termination detection
should eventually occur provided that it remains enabled. Indeed, the formula
$\WF_v(A)$ is defined in \tlaplus as
\[
  \Box\big((\Box\,\ENABLED\,\an{A}{v}) \implies \Diamond\an{A}{v}\big)
\]
which asserts that the action $\an{A}{v}$, defined as $A \land v' \neq v$, must
eventually occur if it ever remains enabled. Enabledness
of an action is defined as
\[
  \ENABLED\ A\ \deq\ \exists v': A
\]
where $v$ is the tuple of variables that have (free) primed occurrences in $A$.

One may consider assuming stronger fairness conditions for this specification,
such as that any pending message should eventually be received. However, weak
fairness of the action $DetectTermination$ is all that is required to ensure
that global termination will be detected.

\section{Verification by Model Checking and Theorem Proving}
\label{sec:atd-verification}

Module $AsynchronousTerminationDetection$ provides a high-level specification of
termination detection; it does not describe a mechanism for solving the problem.
Nevertheless, we can already use the \tlaplus tools in order to verify some
correctness properties, including type correctness, safety, and liveness.

The safety property asserts that termination is not detected unless the system
has indeed terminated, while liveness asserts that termination will eventually
be detected. These properties can be expressed as the temporal formulas
\begin{align*}
    Safe\ &\deq\ \alw SafeInv, \text{ where } SafeInv\ \deq\ termDetect \implies terminated,\\
    Live\ &\deq\ terminated \leadsto termDetect.
\end{align*}
The formula $F \leadsto G$ is shorthand for $\Box(F \implies \Diamond G)$, and it
asserts that whenever~$F$ is true, $G$ must eventually become true. Correctness
of the specification means proving the theorems that $Spec$ implies the above
properties. We may also want to verify that once the system has terminated, it
will remain quiescent, expressed as 
\[
  Quiescence\ \deq\ \alw(terminated \implies \alw terminated).
\]
Note that the two properties $Safe$ and $Quiescence$ imply the derived property
\[
  \alw(termDetect \implies \alw terminated)
\]
asserting that once termination has been detected, the system will remain
globally inactive.

\subsection{Finite-State Model Checking Using \tlc}
\label{sec:atd-mc}

\tlc{} is an explicit-state model checker for checking safety and liveness of
finite instances of \tlaplus specifications~\cite{yu:model-checking}. In order to
describe the finite instance and to indicate the properties to be checked, \tlc{}
requires, in addition to the \tlaplus{} specification, a \emph{configuration
  file}. In our example, we have to provide a value for the constant parameter
$N$ representing the number of nodes. For example, the configuration file in
Figure~\ref{fig:tlc-config} instructs~\tlc{} to check the invariants $TypeOK$
and $Safe$ against the instance of four nodes.
\tlc{} is integrated into both IDEs for \tlaplus{}, the \tlaplus{} Toolbox and
the Visual Studio code extension. Both IDEs generate \tlc{} configuration files
in the background. 

\begin{figure}[tb]
  \begin{minipage}{.3\textwidth}
    \begin{verbatim}
  INIT Init
  NEXT Next      \end{verbatim}
  \end{minipage}
  \begin{minipage}{.6\textwidth}
      \begin{verbatim}
  CONSTANT   N = 4
  INVARIANTS TypeOK Safe   \end{verbatim}
  \end{minipage}
  \caption{A \tlc{} configuration file: It fixes $N$ as 4, the initial and transition
  predicates as $\mathit{Init}$ and $\mathit{Next}$, and the invariants to be verified
  as~$\mathit{TypeOK}$ and $\mathit{Safe}$.}\label{fig:tlc-config}
\end{figure}
  
Since in our specification, nodes may send arbitrarily many messages, the state
space is infinite even for a fixed number of nodes. We therefore add a state
constraint bounding the number of pending messages to $K$ messages per
node.\footnote{More precisely, \tlc will not compute any successors of states
  at which the constraint does not hold.}
\tlc{} quickly verifies all four properties introduced earlier and reports that
the system has 4,097 distinct reachable states for $N=4$ and $K=3$. For $N=6$
and $K=3$, we obtain 262,145 states, illustrating the well-known problem of
state-space explosion.

It is instructive to observe what happens when a specification contains an
error. For example, let us assume that we forgot the conjunct $active[i]$ in the
definition of action $SendMsg(i)$. Running \tlc on the modified specification
indicates that the invariant $Safe$ is violated.\footnote{Property $Quiescence$
  is also violated, but invariant violations are found earlier.}
\tlc produces a counter-example containing a step where all processes are
inactive and termination has been detected, but where a message is sent. In the
successor state, $terminated$ is therefore false, while the variable
$termDetect$ is still true, violating the asserted invariant.

Besides checking the invariants $TypeOK$ and $Safe$, \tlc can also check
properties of specifications expressed as temporal formulas, including
$Quiescence$ and $Live$. For the latter, it is important to include the fairness
assumption in the specification of the instance to be verified by replacing the
\verb|INIT| and \verb|Next| entries in the configuration file by
\verb|SPECIFICATION Spec|.

\subsection{Bounded Model Checking with Apalache}\label{sec:atd-apalache}

\begin{figure}[tb]
  \begin{nomodule}
    \CONSTANT\\
    \hspace*{1em}\verb|\* @type: Int;|\\
    \hspace*{1em}$N$\\
    \VARIABLES\\
    \hspace*{1em}\verb|\* @type: Int -> Bool;|\\
    \hspace*{1em}$active$,\\
    \hspace*{1em}\verb|\* @type: Int -> Int;|\\
    \hspace*{1em}$pending$,\\
    \hspace*{1em}\verb|\* @type: Bool;|\\
    \hspace*{1em}$termDetect$\\
  \end{nomodule}
  \caption{Type annotations for \apalache.}
  \label{fig:ring-tla-apalache}
\end{figure}

\apalache~\cite{konnov:apalache} is a symbolic model checker that leverages the
SMT (satisfiability modulo theories) solver Z3~\cite{DeMouraB08} for
checking~\tlaplus{} executions of bounded lengths as well as proving inductive
invariants. Similar to \tlc{}, \apalache{}
can check instances of specifications where the size of
data structures remains bounded. Reusing the configuration file shown in
Fig.~\ref{fig:tlc-config}, \apalache{} does not actually require a bound $K$ for
the number of pending messages, as it can reason about unbounded integers.


Whereas~\tlc{} enumerates the reachable states one by one, \apalache{} encodes
bounded symbolic executions as a set of constraints, which are either proven to
be unsatisfiable or solved by an SMT solver.
\apalache{} uses the standard
many-sorted first-order logic of SMT, and it infers types of
expressions in a~\tlaplus{} specification based on an annotation of constants
and variables with types, as shown in
Fig.~\ref{fig:ring-tla-apalache}. Its type checker ensures that these
annotations are consistent with the use of the constants and variables in the
expressions that appear in the specification.

Similar to the foundational paper on bounded model checking~\cite{BiereCCZ99},
we write
\[
    Spec \models_k \alw Inv
\]
to denote that a state formula
$Inv$ holds true in all states of all bounded executions of specification
$Spec$ that perform at most~$k$ transitions.
By abuse of notation, we also
allow $Inv$ to be an action formula, which must then be true for all pairs of
subsequent states within these bounded executions, and we call $Inv$ an
\emph{action invariant}.

\begin{figure}[tb]
  \begin{subfigure}[b]{0.48\linewidth}
    \centering
    \includegraphics[width=\textwidth]{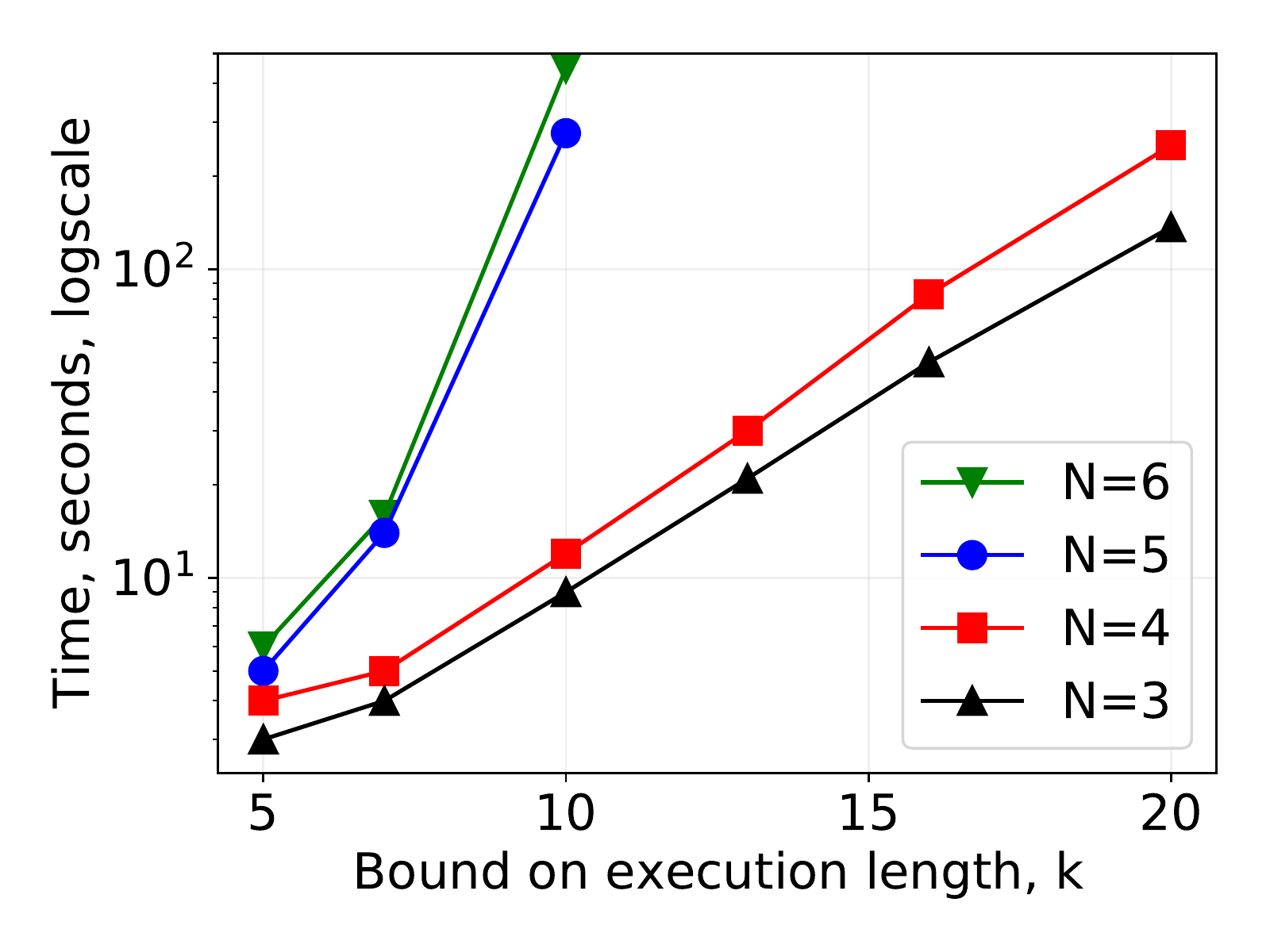}
    \caption{Bounded model checking.}
    \label{fig:apalache-atd:a}
  \end{subfigure}
  \hfill
  \begin{subfigure}[b]{0.48\linewidth}
    \centering
    \includegraphics[width=\textwidth]{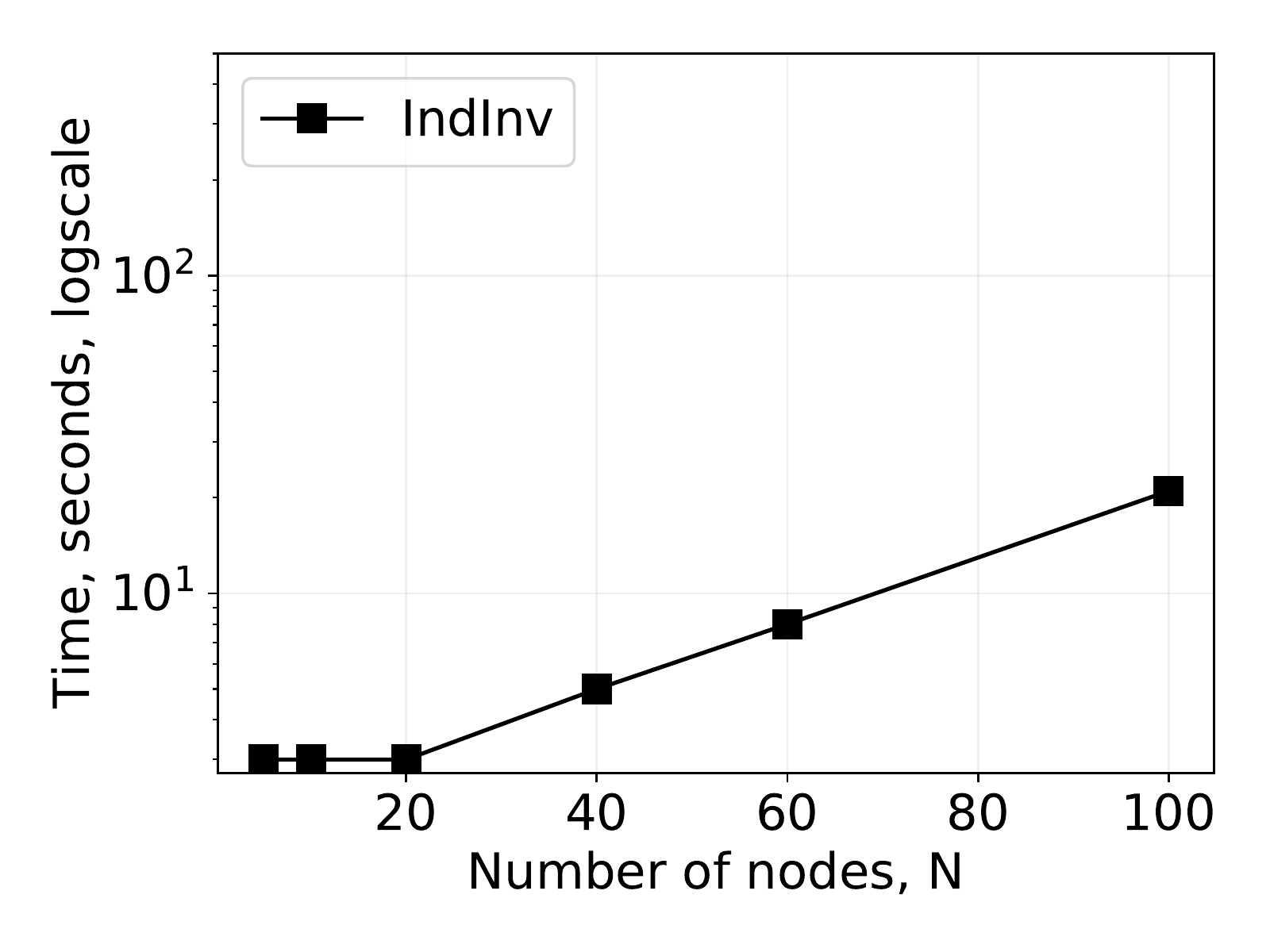}
    \caption{Inductive invariant checking.}
    \label{fig:apalache-atd:b}
  \end{subfigure}
  \caption{Checking $TypeOK \land SafeInv$ using \apalache.}
  \label{fig:apalache-atd}
\end{figure}

\apalache{}
can check that the state invariants~$TypeOK$
and~$SafeInv$ hold of an instance of our specification where $N=4$
and $k \le 10$, in less than a minute.
Figure~\ref{fig:apalache-atd:a} shows the performance of~\apalache{} when
checking the combined invariant $TypeOK \land SafeInv$, for $N \in 3\,..\,6$ and
execution length~$k$ up to 20 steps. As can be seen, \apalache{} suffers from
considerable slowdown when the parameters $N$ and $k$ are increased. This
is caused by combinatorial explosion of the underlying SMT problem, similar
to state explosion of state enumeration in~\tlc{}.


We can also direct \apalache{} to check that the invariant
\[
  IndInv \deq TypeOK \land SafeInv
\]
is inductive and therefore holds for executions of arbitrary length. Informally,
this means that $IndInv$ holds true for the initial states specified
with $Init$, and that it is preserved by all transitions specified with
$Next$. 

%
Using the notation introduced above, we frame these two properties as the
\apalache{} queries \eqref{eq:IndInvInit-apa} and \eqref{eq:IntInvNext-apa}
below. The first query establishes that $IndInv$ is a state invariant of the
original specification for all executions of length~0, and therefore it must
hold in all states that satisfy $Init$. The second query confirms that $IndInv$
is a state invariant for all executions of length~1 when using $IndInv$ as the
initialization predicate instead of $Init$. Therefore, any step of an execution
starting in a state in which $IndInv$ holds preserves the invariant.
\begin{align}
  Init \land \alw\sq{Next}{vars}&\models_0 \alw IndInv
  \label{eq:IndInvInit-apa}\\
  IndInv \land \alw\sq{Next}{vars}&\models_1 \alw IndInv
  \label{eq:IntInvNext-apa}
\end{align}

Both of these runs take only a second on a standard laptop for $N=4$. In fact,
we can show the inductiveness of $\mathit{IndInv}$ for $N=100$ in 20 seconds.
Figure~\ref{fig:apalache-atd:b} shows the performance of~\apalache{} when
checking property~(\ref{eq:IntInvNext-apa}) for various values of~$N$. Comparing
Figs.~\ref{fig:apalache-atd:a} and~\ref{fig:apalache-atd:b}, it becomes apparent
that we can prove inductive invariants of instances of specifications that have
astronomically larger state spaces than those for which standard bounded model
checking is feasible. Moreover, inductive
invariants guarantee that the property holds for executions of arbitrary length.
However, as we will see in Sect.~\ref{sec:ewd998-analysis}, finding useful
inductive invariants is not always as easy as it was in this example.

We cannot directly check the property~$\mathit{Quiescence}$ with~\apalache{},
as it is written as a temporal property. One way of doing so would be to
introduce a history variable that records the sequence of states seen so far and
formulate the property as a state invariant over this history. \apalache{}
offers support for this approach in the form of checking so-called trace
invariants.
%

%
However, we can avoid the encoding as a trace invariant by observing that the
property $Quiescence$ can be verified by checking that the action invariant
$terminated \implies terminated'$ holds for all transitions taken from the
reachable states. Since we have shown that $\mathit{IndInv}$ is an inductive
invariant, it suffices to use \apalache to check the following:
\begin{equation}
  IndInv \land \alw\sq{Next}{vars}
  \models_1 \alw(terminated \implies terminated')
    \label{eq:ind-term-apa}
\end{equation}
It takes~\apalache{} 3 seconds and 11 seconds to show that
property~(\ref{eq:ind-term-apa}) holds for $N=4$ and $N=100$, respectively.

\subsection{Theorem Proving Using \tlaps}\label{sec:atd-prove}

Model checking is invaluable for finding errors, and the
counter-examples computed in the case of property violations help designers
understand their root cause. However, it is restricted to the verification of finite
instances, and it suffers from combinatorial explosion. The \tlaplus Proof
System (\tlaps) \cite{cousineau:tla-proofs} can be used to prove properties of
arbitrary instances of a specification. The effort is independent of the size of
the state space, but it requires the user to write a proof, which is then checked
by the system.

\tlaps does not implement a foundational proof calculus for \tlaplus, but relies
on automatic back-end provers to establish individual proof steps.
Correspondingly, \tlaplus proofs are written as a collection of steps that
together entail the overall theorem. A proof step may be discharged directly by
a back-end, or it may recursively be proved as the consequence of lower-level
steps, leading to a hierarchical proof format~\cite{lamport:howtoprove}. The
proof of an inductive invariant such as $TypeOK$ is written as follows.
\[\begin{noj}
  \THEOREM\ TypeCorrect\ \deq\ Spec \implies \alw TypeOK\\
  \ps{1}{1.}\ Init \implies TypeOK\\
  \ps{1}{2.}\ TypeOK \land \sq{Next}{vars} \implies TypeOK'\\
  \ps{1}{3.}\ \QED\quad \BY\ \ps{1}{1},\,\ps{1}{2},\,\PTL\ \DEF\ Spec
\end{noj}\]
It consists of three top-level steps
$\ps{1}{1}$\,--\,$\ps{1}{3}$.\footnote{Steps are named $\ps{l}{n}$ where $l$
  indicates the nesting level of the step and $n$ is arbitrary.} The first two
steps assert that the initial condition implies the invariant, and that the
invariant is preserved by any transition allowed by $\sq{Next}{vars}$. The final
$\QED$ step corresponds to proving the theorem, assuming the preceding steps
could be proved. The $\BY$ clause directs \tlaps to check the proof of this step by invoking
the $\PTL$ back-end (for ``propositional temporal logic''), assuming steps
$\ps{1}{1}$ and $\ps{1}{2}$, and expanding the definition of formula $Spec$.
$\PTL$ can discharge this proof obligation, which essentially corresponds to an
application of the proof rule
\[
  \begin{array}{@{}c}
    I \implies J \qquad J \land \sq{N}{v} \implies J'\\
    \hline
    I \land \alw\sq{N}{v} \implies \alw J
  \end{array}
\]
of temporal logic. By relying on the $\PTL$ back-end, which implements a
decision procedure for linear-time temporal logic, the user does not have to
indicate a specific proof rule, and \tlaps can discharge more complex steps
involving temporal logic.

In order to complete the proof of the theorem, we must provide proofs for steps
$\ps{1}{1}$ and $\ps{1}{2}$. These steps assert (the validity of) a state and an
action formula, respectively, and therefore do not require temporal
logic.\footnote{\tlaps replaces primed variables occurring in action formulas by
  fresh variables, unrelated to their unprimed counterparts.} The first step can
be proved by invoking the assumption $NAssumption$ and expanding the definitions
of $Init$, $TypeOK$, and the defined operators used therein. The second step
requires some more interaction and is decomposed into one sub-proof per disjunct
in the definition of $Next$. The \tlaplus Toolbox provides an assistant for such
syntactic decompositions. The proofs of the invariant $Safe$ and of the safety
property $Quiescence$ are similar, but they also use theorem $TypeCorrect$ in
order to introduce predicate $TypeOK$ as an assumption.

The proof of the liveness property makes use of the fairness hypothesis that
appears as part of the specification. Since the fairness formula is defined in
terms of $\ENABLED$ (cf.\ Sect.~\ref{sec:td}), we first prove a lemma that
reduces the relevant enabledness condition to a simple state predicate.
\[\begin{noj}
  \LEMMA\ EnabledDT\ \deq\\
  \quad\begin{noj2}
    \ASSUME & TypeOK\\
    \PROVE  & (\ENABLED~\an{DetectTermination}{vars})\\
            & \biimplies (terminated \land \lnot termDetect)
  \end{noj2}
\end{noj}\]
The proof of this lemma makes use of specific directives for reasoning about
$\ENABLED$ provided by \tlaps. The proof of the liveness property is then finished
in a few lines of interaction:
\[\begin{noj}
  \THEOREM\ Liveness\ \deq\ Spec \implies Live\\
  \begin{noj2}
    \ps{1}{.}  & \DEFINE\ P\ \deq\ terminated \land \lnot termDetect\\
    \ps{1}{1.} & TypeOK \land P \land \sq{Next}{vars} \implies P' \lor termDetect'\\
    \ps{1}{2.} & TypeOK \land P \land \an{DetectTermination}{vars} \implies termDetect'\\
    \ps{1}{3.} & TypeOK \land P \implies \ENABLED~\an{DetectTermination}{vars}\\
    \ps{1}{4.} & \QED\quad \BY\ \ps{1}{1},\ \ps{1}{2},\ \ps{1}{3},\
    TypeCorrect,\ \PTL\ \DEF\ Spec,\ Live
  \end{noj2}
\end{noj}\]
The first two steps are proved by expanding the required definitions, the third
step is a consequence of lemma $EnabledDT$, and the final step follows from the
preceding ones, and theorem $TypeCorrect$, by propositional temporal reasoning. 
More complex cases would typically require induction over a well-founded ordering, 
which is supported by a standard library of \tlaplus lemmas.

\section{Safra's Algorithm for Termination Detection}
\label{sec:ewd998}

In his note EWD\,998~\cite{dijkstra:ewd998}, Dijkstra describes an algorithm due
to Safra for detecting termination on a ring of processes. Safra's algorithm
extends a simpler algorithm due to Dijkstra, described in note
EWD\,840~\cite{dijkstra:ewd840}, which assumes that message passing between
processes is instantaneous. In both algorithms, node~$0$ plays the role of a
master node that will detect global termination.

In addition to the activation status and message counter represented by the
variables $active$ and $pending$ introduced for the state machine of
Sect.~\ref{sec:td}, each node now has a color (white or black) and maintains an
integer counter that represents the difference between the numbers of messages
it has sent and received during the execution. In addition, a token circulates on the
ring whose attributes are its color $token.c$, its position $token.p$ (i.e., the
number of the node that the token is currently at), and an integer counter
$token.q$ that represents the sum of the counter values of the nodes visited so
far. Safra's algorithm does not have the variable $termDetect$.

Informally, the algorithm relies on the idea that when the system has globaly
terminated, each node is locally inactive and the sum of the differences between
sent and received messages is $0$. By visiting each node, checking its
activation status and accumulating the counter values, the token reports these
conditions to the master node. Colors are used in order to rule out false
positives: nodes color themselves black at message reception (i.e., when they
may have become active again), and the token becomes black when it passes a
black node. A round in which a black token returns to node $0$ is deemed
inconclusive.

Formally, the algorithm is described by a state machine. The initial values of
each node's activation status and color are arbitrary, all node counters are
$0$, and there are no pending messages. The token is initially black (ensuring
that it will perform at least one full round of the ring), and its initial counter
value is $0$. The token can be located at any of the nodes. The transitions of
the algorithm are as follows.

\begin{description}
\item[$InitiateProbe.$] Node~$0$ may initiate a new round of the token when it
  holds the token (i.e., $token.p=0$) and when the previous round did not detect
  termination: either the token is black, node~$0$ is black, or the sum of the
  counter of node $0$ and $token.q$ is positive. Node~$0$ transfers a fresh white
  token to its neighbor node (i.e., $token.p = N-1$, $token.c = white$, and
  $token.q = 0$) and repaints itself white.
\item[$PassToken.$] A non-zero node~$i$ may pass the token to its neighbor when
  it holds the token (i.e., $token.p=i$) and when it is inactive. After the
  transition, $token.p$ will be $i-1$, $token.q$ is augmented by node~$i$'s counter
  value, and the token will be black if node~$i$ is black or if it was already
  black, whereas node~$i$ becomes white.
\item[$SendMsg.$] This action is similar to the corresponding action of the
  abstract state machine, except that the sender's counter value is incremented
  by~$1$.
\item[$RcvMsg.$] Again, this action is similar to the corresponding action of
  the abstract state machine. However, the receiver's counter value is
  decremented by~$1$, and the receiver becomes black.
\item[$Terminate.$] This action is analogous to the corresponding action of the
  abstract state machine.
\end{description}

Node~$0$ declares global termination when it is inactive, has no pending
messages, holds a white token, its color is white, and the sum of $token.q$ and
its own counter value is $0$. A \tlaplus specification of the algorithm is
available online~\cite{MerzKuppeKonnov2021}.

\section{Analyzing Safra's Algorithm}
\label{sec:ewd998-analysis}

As for the high-level specification of Section~\ref{sec:td}, we use the model
checkers \tlc and \apalache to gain confidence in the correctness of the
algorithm by checking its properties for small instances, and then start writing
a full correctness proof using \tlaps. Instead of rechecking the elementary
correctness properties introduced in Section~\ref{sec:atd-verification}, an
attractive way of verifying the correctness of the algorithm in \tlaplus is to
formally relate it to the high-level algorithm by establishing refinement.

\subsection{Model Checking Correctness Properties}
\label{sec:ewd998-mc}

We start by writing a type-correctness invariant for the \tlaplus specification
of Safra's algorithm and define the predicate of termination detection as
follows:
\[
  termDetect\ \deq\
  \begin{conj}
    token.p = 0 \land token.c = \str{white}\\
    color[0] = \str{white} \land \lnot active[0] \land pending[0] = 0\\
    token.q + counter[0] = 0
  \end{conj}
\]
We use \tlc to verify type correctness, as well as the properties $Safe$,
$Live$, and $Quiescence$ introduced in Sect.~\ref{sec:atd-verification}, for
fixed values of $N$. Again, we need a state constraint for bounding the number
$K$ of pending messages at each node, but also the maximum counter values $C$
and token counter $Q$. The state space of this specification is significantly
larger than that of the higher-level model: fixing $N=3$, $K=C=3$ and $Q=9$,
\tlc finds 1.3~million distinct states and requires 42 seconds on a
desktop-class machine. For $N=4$ and the same bounds for the other parameters,
we obtain 219~million distinct states, and \tlc requires about 50 minutes. While
\tlc scales to multiple cores and can, e.g., verify safety of EWD998 for $N=4$
on a machine with 32 cores and 64 GB of memory in around 10 minutes, it would be
hopeless to model check the specification for larger values such as $N=6$ in
reasonable time.

Experience indicates that by exhaustively checking all reachable states,
including corner cases that would arise very rarely in actual executions of the
algorithm, model checking finds errors in small instances of a specification.
Choosing suitable parameter values is a matter of engineering judgment. For
example, an error introduced into the definition of action $PassToken$ such that
the token adopts the color of the visited node, independently of the current
token color, is found by \tlc when $N\geq4$, but the definition is correct for
$N\leq3$.

If exhaustive model-checking is infeasible due to the size of the state space,
\tlc can verify safety and liveness properties on randomly generated behaviors.
Since randomized state exploration has no notion of state-space coverage, \tlc
runs until either it finds a violation, is manually stopped, or up to a given
resource limit. Because randomized state exploration is an embarrassingly
parallel problem, it scales with the number of available cores.

To illustrate the effectiveness of randomized state exploration with \tlc, we
separately introduced six bugs in the specification of Safra's algorithm that we
observed while teaching \tlaplus classes. Beyond the previously mentioned bug of
not taking into account the token color when non-initiator nodes pass the token,
we initialize the token to white, allow an active node to pass the token, omit
to whiten a node that passes the token, or prevent the initiator from initiating
a new token round when its color is black. Half of these bugs violate property
$Inv$, while the other half violate $TD!Spec$ (see below). With the parameter
value fixed to $N=7$ and no state constraint, randomized state exploration, with
a resource limit of 2 to 3 seconds and for behaviors of length up to 100 states,
finds violations of $Inv$ and $TD!Spec$ in the majority of runs for any of the
six bugs. Thus, the minimal resource usage makes it feasible, and the high
likelihood of findings bugs makes it desirable to automatically run randomized
state exploration repeatedly in the background while writing specifications.
Users can run randomized state exploration with the Visual Studio Code Extension
whenever its editors are saved. Once a spec has matured and randomized state
exploration stops finding bugs, users can switch to exhaustive model-checking.

For more complex specifications, the likelihood of finding bugs with repeated, brief
randomized state exploration can usually be increased further.
If a candidate inductive invariant $IndInv$ is known, using it as the initial
condition makes \tlc explore states that are located at arbitrary depths in the
state space. Should the
set of all states defined by $IndInv$ be too large or even infinite, \tlc can
randomly select a subset from that set with the help of operators defined in the
standard $Randomization$ module. This technique is described in more detail as
part of a note on validating candidates for inductive invariants with
\tlc~\cite{Lamport2018UsingTT}.

\apalache again is particularly useful when it comes to checking inductive
invariants. Dijkstra's note~\cite{dijkstra:ewd998} introduces the following
inductive invariant $Inv$ (written in \tlaplus):
\[
  \begin{noj2}
    Sum(f,S) & \deq\ FoldFunctionOnSet(+, 0, f, S)\\
    Rng(a,b) & \deq\ \{i \in Node: a \leq i \land i \leq b \}\\
    Inv & \deq\ 
    \begin{conj}
      Sum(pending, Node) = Sum(counter, Node)\\
      \begin{disj}
        \begin{conj}
          \A i \in Rng(token.p+1, N-1): active[i] = \FALSE\\
          token.q = Sum(counter, Rng(token.p+1, N-1))
        \end{conj}\\
        Sum(counter, Rng(0, token.p)) + token.q > 0\\
        \E i \in Rng(0, token.p) : color[i] = \str{black}\\
        token.c = \str{black}
      \end{disj}
    \end{conj}
  \end{noj2}
\]
The expression $Sum(f,S)$ represents the sum of $f[x]$ for all $x \in S$; it is
defined in terms of the operator $FoldFunctionOnSet$ from the \tlaplus
Community Modules~\cite{A_Kuppe_TLA_CommunityModules}, a
collection of useful libraries for use with \tlaplus.

As explained in Sect.~\ref{sec:atd-apalache}, we can use \apalache{} to check that
$Inv$ is indeed an inductive invariant for finite instances of the
specification. For $N=4$, this takes 11 seconds. It is
not hard to see that $Inv$, together with predicate $termDetect$, implies
$terminated$. Indeed, $termDetect$ implies that the three final disjuncts of the
invariant are false, hence the first disjunct must be true. Thus, all nodes are
inactive, and $token.q$ equals the sum of the counter values of nodes
$1\,..\,N-1$. By $termDetect$, it follows that the sum of the counter values of
all nodes must be $0$, and by the first conjunct of $Inv$ it follows that there
are no pending messages at any node, hence $terminated$ is true.

\subsection{Safra's Algorithm Implements Termination Detection}
\label{sec:ewd998-refines}

Instead of verifying the \tlaplus specification of Safra's algorithm against
correctness properties such as $Safe$ and $Live$, we can show that it implements
the high-level state machine of Sect.~\ref{sec:td}. It then follows that the
properties verified for that state machine are ``inherited'' by the low-level
specification. Since in \tlaplus, refinement is implication, this assertion can
be stated by inserting the following lines in the module representing Safra's
algorithm:
\[\begin{noj}
    TD\ \deq\ \INSTANCE\ AsynchronousTerminationDetection\\
    \THEOREM\ Refinement\ \deq\ Spec \implies TD!Spec
  \end{noj}\]
The first line declares an instance $TD$ of the high-level specification in
which the constant parameter $N$ and the variable parameters $active$, $pending$
and $termDetect$ are instantiated by the expressions of the same name in the
specification of Safra's algorithm.\footnote{In general, \INSTANCE{} allows
  constant and variable parameters to be instantiated by expressions defined in
  terms of the operators defined in the current context.} Theorem
$Refinement$ asserts that every run of $Spec$ (the specification of Safra's
algorithm) also satisfies $TD!Spec$, defined in Fig.~\ref{fig:atd-tla}.

\tlc can reasonably verify refinement for values of $N<5$, and with a similar state
constraint as before, simply by indicating $TD!Spec$ as the temporal property
to be checked. \apalache cannot handle the fairness condition that is part of
$TD!Spec$, but it can verify initialization and step simulation.
Technically, this is done by checking one state invariant and one action
invariant with~\apalache{}:
\begin{align}
    Init \land \alw\sq{Next}{vars}&\models_0 \alw TD!Init
        \label{eq:ref-init}
        \\
    TypeOK \land Inv \land \alw\sq{Next}{vars}&\models_1 \alw \sq{TD!Next}{TD!vars}
        \label{eq:ref-next}
\end{align}

Checking condition~(\ref{eq:ref-init}) with the~\apalache{} machinery is
equivalent to showing $\mathit{Init} \Rightarrow TD!Init$, which is needed to
show the initialization property of refinement.  To prove step simulation, we
have to show that any transition described by $\sq{Next}{vars}$, starting in
any reachable state of the low-level specification, simulates a high-level
transition according to $\sq{TD!Next}{TD!vars}$.  As expressed in condition
(\ref{eq:ref-next}), the previously established inductive invariants $TypeOK$
and $Inv$ are sufficient for proving step simulation.

The \tlaplus specification of Safra's algorithm as described in
Sect.~\ref{sec:ewd998} can be refined further by introducing explicit message
channels and node addresses. We refer readers to the modules $EWD998Chan$ and
$EWD998ChanID$~\cite{MerzKuppeKonnov2021} that contain corresponding \tlaplus
specifications for which refinement can be checked using the \tlaplus tools.

\subsection{Proving Correctness Using \tlaps}
\label{sec:ewd998-tlaps}

After gaining confidence in the validity of our specification of Safra's
algorithm, we again use \tlaps for proving its correctness for arbitrary
instances. The type correctness proof is quite similar to that of the high-level
specification described in Section~\ref{sec:atd-prove}. We also used \tlaps for
proving the inductive invariant $Inv$ introduced in Sect.~\ref{sec:ewd998-mc}.
However, at the time of writing, no theorem libraries exist in \tlaps for operators
such as $FoldFunctionOnSet$, and we therefore stated elementary properties
without proof, and specialized them for the derived operator $Sum$ such as
\[\begin{noj}
  \LEMMA\ SumIterate\ \deq\\
  \quad\begin{noj2}
    \ASSUME & \NEW\ fun \in [Node \fun Int], \\
            & \NEW\ inds \in \SUBSET\ Node,\ \NEW\ e \in inds\\
    \PROVE  & Sum(fun, inds) = fun[e] + Sum(fun, inds \setminus \{e\})
  \end{noj2}
\end{noj}\]

The invariance proof itself was written in one person-day and required about 230
lines in the proof language of \tlaplus. The hierarchical proof style helps to
focus on individual steps without having to remember the overall proof.

Based on the inductive invariant and the following lemma
\[\begin{noj}
  \LEMMA\ Safety\ \deq\ 
  TypeOK \land Inv \land termDetect \implies Termination
\end{noj}\]
that formalizes the argument presented at the end of Sect.~\ref{sec:ewd998-mc},
it is not hard to use \tlaps for proving the safety part of the refinement
relation, i.e.\ the implications expressed by properties \eqref{eq:ref-init} and
\eqref{eq:ref-next}, for arbitrary instances of the specification. The proofs of
the lemma and the safety part of the refinement theorem require about 110 lines
of \tlaplus proof and were written in half a person-day.

In order to prove liveness of the algorithm, we first reduce the enabledness
condition of the action for which fairness is assumed, to a simple state
predicate, as we did in Sect.~\ref{sec:atd-prove}. We must then show that any
state satisfying predicate $terminated$ must be followed by one where
$termDetect$ holds. Relying on the already proved invariants,
we set up a proof by contradiction and define
\[
  BSpec\ \deq\ \alw TypeOK \land \alw Inv \land \alw\lnot termDetect
               \land \alw\sq{Next}{vars} \land \WF_{vars}(System)
\]
where the last conjunct corresponds to the fairness assumption of the specification,
in which $System$ represents the disjunction of the token-passing transitions.
Informally, detecting termination may require three rounds of the token:
\begin{enumerate}
\item The first round brings the token back to node $0$, while $terminated$
  remains true.
\item After the second round, the token is back at node $0$, all nodes are white
  (since no messages could be received), and $terminated$ is still true.
\item At the end of the third round, the same conditions hold, and additionally
  the token is white and the token counter holds the sum of the counters of the
  nodes in the interval $1\,..\,N-1$.
\end{enumerate}

We prove a lemma corresponding to each of the rounds. For example, for the first
round we assert
\[
  \LEMMA\ Round1\ \deq\ BSpec \implies (terminated \leadsto (terminated \land token.p=0))
\]
The proof proceeds by induction on the current position~$k$ of the node: if
$k=0$, the assertion is trivial. For the case $k+1$, any step of the system
either leaves the token at node $k+1$ or brings it to node~$k$. Moreover, the
token-passing transition from node $k+1$ is enabled, and it takes the token to node~$k$.

The statements and proofs of the two other lemmas are similar. Finally, the
post-condition of the third round, together with $Inv$, implies $termDetect$,
concluding the proof. As a corollary, we can finish the refinement proof: we
showed in Sect.~\ref{sec:atd-prove} that the enabledness condition of
$TD.DetectTermination$ corresponds to the conjunction $terminated \land \lnot
termDetect$, and we just proved that this predicate cannot hold forever,
implying the fairness condition. The \tlaplus proof takes 245 lines and was
written in less than one person-day, aided by the fact that the three main
lemmas and their proofs are quite similar.

\section{Conclusion}
\label{sec:conclusion}

\tlaplus{} is a language for the formal and unambiguous description of
algorithms and systems. In this paper, we presented the three main tools for
verifying properties of \tlaplus specifications: the explicit-state model checker \tlc{},
the symbolic model checker \apalache{}, and the interactive proof assistant
\tlaps{}, at the hand of a formal specification of a non-trivial distributed
algorithm. The ProB model checker can also verify \tlaplus specifications
through a translation to B~\cite{hansen:translating}, but we did not evaluate
its use on the case study presented here. The three tools that we considered
have complementary strengths and weaknesses. They offer various degrees of proof
power in exchange for manual effort or computational resources. Whereas~\tlc{}
is essentially a push-button tool, \tlaps{} requires significant user effort
for inventing a proof. Likewise, \apalache{} can sometimes prove invariants of
specifications that have significantly larger (or even infinite) state spaces
than~\tlc{} can handle, but it requires human ingenuity to find an inductive invariant.
Hence, we advocate the following basic workflow.

Since errors in specifications are usually found for small instances, it is
easiest to start by using \tlc for checking basic invariant properties. The same
properties can also be checked using \apalache for short execution prefixes.
However, \apalache really shines for checking inductive (state and action)
invariants since doing so only requires considering a single transition. Once
reasonable confidence has been obtained for safety properties, liveness
properties can be verified using \tlc. One potential pitfall here is that 
the use of a state constraint
may mask non-progress cycles in which some variable value exceeds the admissible
bounds. Both model checkers suffer from combinatorial explosion;
for \apalache the length of the execution prefixes that need to be examined tends to be the
limiting factor. For very large state spaces, \tlc provides support for
randomized state exploration, which empirically tends to find bugs with good
success when exhaustive exploration fails. When model checking finds no more
errors and even more confidence is required, one can start writing a proof and
check it using \tlaps. In this way, properties of arbitrary instances can be
verified, at the expense of human effort that can become substantial in the
presence of complex formulas.

All three tools accept the same input language~\tlaplus{}.  Whereas
\apalache{} requires writing typing annotations for state variables and certain
operator definitions, these are usually not very difficult to come up with.
Not surprisingly, the tools pose additional
restrictions on the input specifications. For example:

\begin{enumerate}
\item \tlc{} rejects action formulas of the form $x' \in S$ when $S$ is an infinite
  set, and no unique value has been determined for the variable $x'$ by an earlier
  conjunct: in this case, the formula would require \tlc{} to enumerate an infinite set.
  However, the user can override operator definitions such as $S$ during
  model-checking without modifying the original specification.

\item \apalache{} has recently dropped support for recursive operators and
  functions in favor of fold operations, e.g., $\mathit{FoldSet}$ and
  $\mathit{FoldFunction}$. Folding a set~$S$ of bounded cardinality, that is,
  $|S| \le n$, for some $n$, needs up to~$n$ iterations, which is easy to encode
  in SMT as opposed to general recursion.

\item \tlaps{} currently does not handle recursive operator definitions, and
  support for proving liveness properties has only recently been added. In practice,
  reasoning about operators that are not supported natively by its backends
  requires well-developed theorem libraries.
\end{enumerate}

Despite these limitations, all three tools share a large common fragment
of~\tlaplus{}. This ability to use different tools for the same specification is
extremely valuable in practice, for example for using the model checkers to
verify a putative inductive invariant in the middle of writing an interactive
proof.

Future work should focus on improving the capabilities of each tool. Although
\tlc is quite mature, it could benefit from better parallelization of liveness
checking. Recent work has focused on improved presentation of counter-examples,
including their visualization and animation~\cite{Schultz2018}, and on randomized
exploration of state spaces. For \apalache, alternative encodings of bounded verification
problems as SMT problems could help with performance degradation when
considering longer execution prefixes. It would also be useful to lift the
current restriction to the verification of safety properties and consider
bounded verification of liveness properties. \tlaps users would appreciate
better automation, such as the tool being able to indicate which operator
definitions should be expanded, better support for higher-order problems, and
for verifying liveness properties. The IDEs could help with using the tools
synergistically, such as starting an exploratory model checking run that
corresponds to a given step in a proof.

Besides the work on verifying properties of formal specifications, there is
interest in relating \tlaplus specifications to system implementations. The two
main lines of research are model-based test case generation from formal
specifications and trace validation, in which the specification serves as a
monitor for supervising an implementation. Davis et al.\ \cite{davis:extreme}
present an interesting real-life case study, in which trace validation was found
to be particularly successful.

\bibliographystyle{splncs04}
\bibliography{paper}

\end{document}